\documentclass[reprint,amsmath,showpacs,amssymb,aip,groupedaddress,pop]{revtex4-1}

\usepackage[english]{babel}
\usepackage{graphicx}
\usepackage{dcolumn}
\usepackage{bm}
\usepackage{hyperref}

\usepackage{color}
\usepackage{enumitem}
\newcommand{\angstrom}{\textup{\AA}}

\begin{document}
\title{Structural, thermodynamic, and transport properties of CH$_2$ plasma in the two-temperature regime}

\author{D.~V.~Knyazev$^{1,2,3}$ and P.~R.~Levashov$^{1,4}$}
\affiliation{$^1$Joint Institute for High Temperatures RAS, Izhorskaya 13 bldg. 2, Moscow 125412, Russia \\ 
$^2$Moscow Institute of Physics and Technology (State University), Institutskiy per. 9, Dolgoprudny, Moscow Region 141700, Russia\\
$^3$State Scientific Center of the Russian Federation -- Institute for Theoretical and Experimental Physics of National Research Centre ``Kurchatov Institute'', Bolshaya Cheremushkinskaya 25, 117218, Moscow, Russia\\
$^4$Tomsk State University, Lenin Prospekt 36, Tomsk 634050, Russia}


\begin{abstract}

This paper covers calculation of radial distribution functions, specific energy and static electrical conductivity of CH$_2$ plasma in the two-temperature regime. The calculation is based on the quantum molecular dynamics, density functional theory and the Kubo-Greenwood formula.

The properties are computed at 5~kK~$\leq T_i\leq T_e\leq40$~kK and $\rho=0.954$~g/cm$^3$ and depend severely on the presence of chemical bonds in the system. Chemical compounds exist at the lowest temperature $T_i=T_e=5$~kK considered; they are destroyed rapidly at the growth of $T_i$ and slower at the increase of $T_e$.

A significant number of bonds are present in the system at 5~kK~$\leq T_i\leq T_e\leq10$~kK. The destruction of bonds correlates with the growth of specific energy and static electrical conductivity under these conditions.

\end{abstract}

\maketitle

\section{Introduction}

Carbon-hydrogen plastics are widely used nowadays in various experiments on the interaction of intense energy fluxes with matter. One of these fruitful applications is described in the paper by Povarnitsyn et al.:\cite{Povarnitsyn:LasPartBeams:2013} a polyethylene film may be used to block the prepulse, and thereby, to improve the contrast of an intense laser pulse.

Two types of prepulse are considered:\cite{Povarnitsyn:LasPartBeams:2013} nanosecond (intensity $I_\mathrm{ns}=10^{13}$~W/cm$^2$; duration $t_\mathrm{ns}=2$~ns) and picosecond ($I_\mathrm{ps}=10^{15}$~W/cm$^2$; $t_\mathrm{ps}=20$~ps) ones . Both prepulses absorbed by the film produce the state of plasma with an electron temperature $T_e$ exceeding an ion temperature $T_i$. The conditions with $T_e>T_i$ are often called a two-temperature ($2T$) regime. The appearance of the $2T$-state may be explained as follows. The absorption of laser radiation by electrons assists the creation of the $2T$-state: the larger laser intensity $I$, the faster $T_e-T_i$ grows. The electron-phonon coupling destroys the $2T$-state: the larger electron-phonon coupling constant $G$, the faster $T_e-T_i$ decreases. Thus if the prepulse intensity $I$ is great enough, the $2T$-state with considerable $T_e-T_i$ may be created.

The action of the prepulse may be described quantitatively using numerical simulation \cite{Povarnitsyn:LasPartBeams:2013, Povarnitsyn:PRB:2015}. Modelling \cite{Povarnitsyn:LasPartBeams:2013} shows, that after a considerable part of the nanosecond prepulse has been absorbed, the following conditions are obtained: relative change of density $\rho/\rho_0=10^{-4}$--$10^{-2}$, $T_i\sim400$~kK, $T_e\sim4\cdot10^3$~kK. A number of matter properties are required to simulate the action of the prepulse. Particularly, an equation of state, a complex dielectric function and a thermal conductivity coefficient should be known. Paper \cite{Povarnitsyn:LasPartBeams:2013} employs rather rough models of matter properties. Therefore, the need for better knowledge of plasma properties arises.

The matter properties should be known for all the states of the system: from the ambient conditions to the extreme parameters specified above. The required properties may be calculated via various techniques, including: the average atom model \cite{Ovechkin:HEDP:2014}, the chemical plasma model \cite{Apfelbaum:CPP:2016} and quantum molecular dynamics (QMD). None of these methods may yield data for all conditions emerging under the action of the prepulse. The QMD technique is a powerful tool for the calculation of properties in the warm dense matter regime (rather high densities and moderate temperatures).

QMD is widely used for the calculation of thermodynamic properties, including equation of state \cite{Wang:PhysPlasmas:2013}, shock Hugoniots \cite{Wang:PRE:2013:2, Minakov:JAP:2014} and melting curves \cite{Minakov:PRB:2015}. A common approach to obtain electronic transport and optical properties from a QMD simulation is to use the Kubo-Greenwood formula (KG). Here the transport properties encompass static electrical conductivity and thermal conductivity, whereas the optical properties include dynamic electrical conductivity, complex dieletric function, complex refraction index and reflectivity. The QMD+KG technique became particularly widespread after the papers \cite{Desjarlais:PRE:2002, Recoules:PRB:2005}. Some of the most recent QMD+KG calculations address transport and optical properties of deuterium \cite{Hu:PhysPlasmas:2015}, berillium \cite{Li:PhysPlasmas:2015}, xenon \cite{Norman:PRE:2015} and copper \cite{Migdal:APA:2016}.

Carbon-hydrogen plasma has also been explored by the QMD technique recently. Pure C$_m$H$_n$ plasma (carbon and hydrogen ions are the only ions present in the system) was considered in papers \cite{Mattsson:PRB:2010, Wang:PhysPlasmas:2011, Lambert:PRE:2012, Hamel:PRB:2012, Chantawansri:JCP:2012, Hu:PRE:2014:1, Danel:PRE:2015}. Other works \cite{Horner:PRB:2010, Magyar:PRB:2015, Huser:PRE:2015, Colin-Lalu:PRE:2015} study the influence of dopants. Transport and optical properties of carbon-hydrogen plasma were investigated in papers \cite{Horner:PRB:2010, Wang:PhysPlasmas:2011, Lambert:PRE:2012, Hu:PRE:2014:1, Huser:PRE:2015}. Some of the cited works are discussed in more detail in our previous work \cite{Knyazev:PhysPlasmas:2015}.

None of the papers mentioned above studies the influence of the $2T$-state on the properties of carbon-hydrogen plasma. The lack of such data was the first reason stimulating this work. In this paper we calculate specific energy and static electrical conductivity of CH$_2$ plasma in the $2T$-state via the QMD+KG technique. The plasma of CH$_2$ composition corresponds to polyethylene heated by laser radiation. In this work the properties are calculated at the normal density of polyethylene $\rho=0.954$~g/cm$^3$ and at temperatures 5~kK~$\leq T_i\leq T_e\leq40$~kK. These conditions correspond to the very beginning of the prepulse action; the temperatures after the prepulse are much larger. However, the beginning of the prepulse action should be also simulated carefully, since the spacial distribution of plasma at the initial stage influences the whole following process dramatically. Thus we have to know the properties of CH$_2$ plasma even for such moderate temperatures.

Properties of CH$_2$ plasma in the one-temperature ($1T$) case $T_i=T_e$ were investigated in our previous work \cite{Knyazev:PhysPlasmas:2015}. The properties were calculated at $\rho=0.954$~g/cm$^3$ and at temperatures 5~kK~$\leq T_i=T_e=T\leq100$~kK. 

The most interesting results obtained in \cite{Knyazev:PhysPlasmas:2015} concern specific heat capacity and static electrical conductivity of CH$_2$ plasma. The specific heat capacity $\mathcal{C}_v(T)$ decreases at 5~kK~$\leq T\leq15$~kK and increases at 15~kK~$\leq T\leq100$~kK. The decrease of $\mathcal{C}_v(T)$ corresponds to the concave shape of the temperature dependence of specific energy $\mathcal{E}(T)$. The temperature dependence of the static electrical conductivity $\sigma_{1_\mathrm{DC}}(T)$ demonstrates step-like behavior: it grows rapidly at 5~kK~$\leq T\leq10$~kK and remains almost constant at 20~kK~$\leq T\leq60$~kK. Similar step-like curves for reflectivity along principal Hugoniots of carbon-hydrogen plastics were obtained in the previous works \cite{Wang:PhysPlasmas:2011, Hu:PRE:2014:1, Huser:PRE:2015}.

The second reason for the current work is the drive to explain the obtained $\mathcal{E}(T)$ and $\sigma_{1_\mathrm{DC}}(T)$ dependences. During a $2T$-calculation one of the temperatures ($T_i$ or $T_e$) is kept fixed while the other one is varied. This helps to understand better the influence of $T_i$ and $T_e$ on the one-temperature $\mathcal{E}(T)$ and $\sigma_{1_\mathrm{DC}}(T)$ dependences.

The structure of our paper is quite straightfoward. Sec.~\ref{Sec:Technique} contains a brief description of the computation method. The technical parameters used during the calculation are available in Sec.~\ref{Sec:Parameters}. The results on $\mathcal{E}$ and $\sigma_{1_\mathrm{DC}}$ of CH$_2$ plasma are presented in Sec.~\ref{Sec:Results}. The discussion of the results based on the investigation of radial distribution functions (RDFs) is also available in Sec.~\ref{Sec:Results}.

\section{Computation technique}
\label{Sec:Technique}

The computation technique is based on quantum molecular dynamics, density functional theory (DFT) in its Kohn-Sham formulation and the Kubo-Greenwood formula. The method of calculation for $1T$ case was described in detail in our previous work \cite{Knyazev:COMMAT:2013} and the papers \cite{Desjarlais:PRE:2002, Recoules:PRB:2005}. An example of $2T$-calculation is present in our previous paper \cite{Knyazev:PhysPlasmas:2014}. Here we will give only a brief overview of the employed technique.

The computation method consists of three main stages: QMD simulation, precise resolution of the band structure and the calculation via the KG formula.

At the first stage $N_C$ atoms of carbon and $N_H=2N_C$ atoms of hydrogen are placed in a supercell with periodic boundary conditions. The total number of atoms $N_\mathrm{at}=N_C+N_H$ may be varied. At the given $N_\mathrm{at}$ the size of the supercell is chosen to yield the correct density $\rho$. Ions are treated classically. The ions of carbon and hydrogen are placed in the random nodes of the auxiliary simple cubic lattice. We have discussed the choice of the initial ionic configuration and performed an overview of the works on this issue previously \cite{Knyazev:PhysPlasmas:2015}. Then the QMD simulation is performed.

The electronic structure is calculated at each QMD step within the Born-Oppenheimer approximation: electrons totally adjust to the current ionic configuration. This calculation is performed within the framework of DFT: the finite-temperature Kohn-Sham equations are solved. The occupation numbers used during their solution are set by the Fermi-Dirac distribution. The latter includes the electron temperature $T_e$; this is how the calculation depends on $T_e$.

The forces acting on each ion from the electrons and other ions are calculated at every step. The Newton equations of motion are solved for the ions using these forces; thus the ionic trajectories are calculated. Additional forces are also acting on the ions from the Nos\'e thermostat. These forces are used to bring the total kinetic energy of ions $E_i^\mathrm{kin}(t)$ to the average value $\frac{3}{2}(N_\mathrm{at}-1)kT_i$ after some period of simulation; here $k$ is the Boltzmann constant. This is how the calculation depends on $T_i$. QMD simulation is performed using the Vienna \textit{ab initio} simulation package (VASP) \cite{Kresse:PRB:1993, Kresse:PRB:1994, Kresse:PRB:1996}.

The ionic trajectories and the temporal dependence of the energy without the kinetic contribution of ions $[E-E_i^\mathrm{kin}](t)$ are obtained during the QMD simulation. The system comes to a two-temperature equilibrium state after some number of QMD steps is performed. In this state equilibrium exists only within the electronic and ionic subsystems separately. The exchange of energy between electrons and ions is absent, since the Born-Oppenheimer approximation is used during the QMD simulation. $[E-E_i^\mathrm{kin}](t)$ fluctuates around its average value in this two-temperature equilibrium state. A significant number of \textit{sequential} QMD steps corresponding to the two-temperature equilibrium state are chosen. $[E-E_i^\mathrm{kin}](t)$ dependence is averaged over these sequential steps; thus the thermodynamic value $[E-E_i^\mathrm{kin}]$ is obtained. If the dependence on time is not mentioned, $[E-E_i^\mathrm{kin}]$ denotes the thermodynamic value of the energy without the kinetic contribution of ions here. The thermodynamic value is then divided by the mass of the supercell; this specific energy is designated by $[\mathcal{E}-\mathcal{E}_i^\mathrm{kin}]$.

The total energy of electrons and ions $E$ may also be calculated. However, these data are not presented in this paper to understand better the temperature dependence of energy. The thermodynamic value of $E_i^\mathrm{kin}$ is $\frac{3}{2}(N_\mathrm{at}-1)kT_i$ because of the interaction with the Nos\'e thermostat. Thus $E_i^\mathrm{kin}$ depends only on $T_i$ in a rather simple way: $E_i^\mathrm{kin}(T_i)\sim T_i$. $[E-E_i^\mathrm{kin}]$ may depend both on $T_i$ and $T_e$ in a complicated manner. The addition of $E_i^\mathrm{kin}(T_i)$ will just obscure the $[E-E_i^\mathrm{kin}](T_i,T_e)$ dependence. If necessary, the total energy may be reconstructed easily.

The \textit{separate} configurations corresponding to the two-temperature equilibrium state are selected for the calculation of static electrical conductivity and optical properties. At the second stage the precise resolution of the band structure is performed for these separate configurations. The same Kohn-Sham equations as during the first stage are solved, though the technical parameters yielding a higher precision may be used. At this stage the Kohn-Sham eigenvalues, corresponding wave functions and occupation numbers are obtained. The precise resolution of the band structure is performed with the VASP package. Then the obtained wave functions are used to calculate the matrix elements of the nabla operator; this is done using the optics.f90 module of the VASP package.

At the third stage the real part of the dynamic electrical conductivity $\sigma_1(\omega)$ is calculated via the KG formula presented in our previous paper \cite{Knyazev:COMMAT:2013}; the formula includes matrix elements of the nabla operator, energy eigenvalues and occupation numbers calculated during the precise resolution of the band structure. We have created a parallel program to perform a calculation according to the KG formula, it uses data obtained by the VASP as input information. $\sigma_{1_j}(\omega)$ is obtained for each of the selected ionic configurations. These $\sigma_{1_j}(\omega)$ curves are then averaged to get the final $\sigma_1(\omega)$. The static electrical conductivity $\sigma_{1_\mathrm{DC}}$ is calculated via an extrapolation of $\sigma_1(\omega)$ to zero frequency. The simple linear extrapolation described in \cite{Knyazev:COMMAT:2013} is used.

A number of \textit{sequential} ionic configurations corresponding to the $2T$ equilibrium stage of the QMD simulation are also used to calculate RDFs. C-C, C-H and H-H RDFs are calculated with the Visual Molecular Dynamics program (VMD) \cite{Humphrey:JMG:1996}.

\section{Technical parameters}
\label{Sec:Parameters}

The QMD simulation was performed with 120 atoms in the computational supercell (40 carbon atoms and 80 hydrogen atoms). At the initial moment the ions were placed in the random nodes of the auxiliary simple cubic lattice (discussed in \cite{Knyazev:PhysPlasmas:2015}). Then 15000 steps of the QMD simulation were performed, one step corresponded to 0.2~fs. Thus the evolution of the system during 3~ps was tracked. The calculation was run in the framework of the local density approximation (LDA) with the Perdew--Zunger parametrization (set by Eqs.~(C3), (C5) and Table~XII, the first column, of paper\cite{Perdew:PRB:1981}). We have applied the pseudopotentials of the projector augmented-wave (PAW \cite{Blochl:PRB:1994, Kresse:PRB:1999}) type both for carbon and hydrogen. The PAW pseudopotential for carbon took 4 electrons into account ($2s^22p^2$); the core radius $r_c$ was equal to $1.5a_B$, here $a_B$ is the Bohr radius. The PAW pseudopotential for hydrogen allowed for 1 electron per atom ($1s^1$), $r_c=1.1a_B$. The QMD simulation was performed with 1 \textbf{k}-point in the Brillouin zone ($\Gamma$-point) and with the energy cut-off $E_\mathrm{cut}=300$~eV. All the bands with occupation numbers larger than $5\times10^{-6}$ were taken into account. The section of the QMD simulation corresponding to 0.5~ps~$\leq t\leq2.5$~ps was used to average the temporal dependence $[E-E_i^\mathrm{kin}](t)$.

We have chosen 15 ionic configurations for the further calculation of $\sigma_{1_\mathrm{DC}}$. The first of these configurations corresponded to $t=0.2$~ps, the time span between the neighboring configurations also was $0.2$~ps. The band structure was calculated one more time for these selected configurations. The exchange-correlation functional, pseudopotential, number of $\textbf{k}$-points and energy cut-off were the same as during the QMD simulation. Additional unoccupied bands were taken into account, they spanned an energy range of 40~eV.

The $\sigma_{1_j}(\omega)$ curves were calculated for the selected ionic configurations at 0.005~eV~$\leq\omega\leq40$~eV with a frequency step 0.005~eV. The $\delta$-function in the KG formula was broadened by the Gaussian\cite{Desjarlais:PRE:2002} function with the standard deviation of 0.2~eV.

The RDFs were calculated for the section of the QMD simulation corresponding to 1.8002~ps~$\leq t\leq2$~ps for the distances 0.05~\angstrom$~\leq r\leq4$~\angstrom\ with the step of $\Delta r=0.1$~\angstrom.

For computational results to be reliable, the convergence by technical parameters should be checked. However, the full investigation of convergence is very time-consuming. In our previous works we have performed full research on convergence for aluminum \cite{Knyazev:COMMAT:2013} and partial---for CH$_2$ plasma \cite{Knyazev:PhysPlasmas:2015}. It was shown \cite{Knyazev:COMMAT:2013}, that the number of atoms $N_\mathrm{at}$ and the number of \textbf{k}-points $N_\mathbf{k}$ during the precise resolution of the band structure contribute to the error of $\sigma_{1_\mathrm{DC}}$ most of all; the effect of these parameters is of the same order of magnitude. The size effects in CH$_2$ plasma were investigated earlier:\cite{Knyazev:PhysPlasmas:2015} the results for $\sigma_{1_\mathrm{DC}}$ were the same within several percents for 120 and 249 atoms in the supercell. In our current paper we use the same moderate $N_\mathrm{at}$ and $N_\mathbf{k}$ values as in \cite{Knyazev:PhysPlasmas:2015}. This introduces a small error to our results, but speeds up computations considerably.

\section{Results}
\label{Sec:Results}

The following types of calculations were performed:
\begin{itemize}[label={--},noitemsep,nolistsep]
\item one-temperature case with $T_i=T_e=T$;
\item $2T$-computations at fixed $T_i$ and varied $T_e$, so that $T_e\geq T_i$; 
\item $2T$-computations at fixed $T_e$ and varied $T_i$, so that $T_i\leq T_e$.
\end{itemize}

The overall range of temperatures under consideration is 5~kK~$\leq T_i\leq T_e\leq 40$~kK. The calculations were performed at fixed $\rho=0.954$~g/cm$^3$. 

These calculations allow us to obtain the dependence of a quantity $f$ both on $T_i$ and $T_e$: $f(T_i,T_e)$. Here $f$ may stand for $[\mathcal{E}-\mathcal{E}_i^\mathrm{kin}]$ or $\sigma_{1_\mathrm{DC}}$. The one-temperature dependences $f(T)|_{T_i=T_e}$ were investigated previously \cite{Knyazev:PhysPlasmas:2015}. In this paper we also present $f(T_e)|_{T_i=\mathrm{const}}$ and $f(T_i)|_{T_e=\mathrm{const}}$. The slope of $f(T_e)|_{T_i=\mathrm{const}}$ equals $\left(\frac{\partial f}{\partial T_e}\right)_{T_i}$; the slope of $f(T_i)|_{T_e=\mathrm{const}}$---$\left(\frac{\partial f}{\partial T_i}\right)_{T_e}$. The slope of $f(T)|_{T_i=T_e}$ may be designated by $\left(\frac{\partial f}{\partial T}\right)_{T_i=T_e}$. Then the following equation is valid:
\begin{equation}
\left(\frac{\partial f}{\partial T}\right)_{T_i=T_e}=\bigg[\left(\frac{\partial f}{\partial T_e}\right)_{T_i}+\left(\frac{\partial f}{\partial T_i}\right)_{T_e}\bigg]_{T_i=T_e}.
\label{Eq:Derivatives}
\end{equation}
If we compare the contributions of $\left(\frac{\partial f}{\partial T_e}\right)_{T_i}$ and $\left(\frac{\partial f}{\partial T_i}\right)_{T_e}$ to $\left(\frac{\partial f}{\partial T}\right)_{T_i=T_e}$ we can understand, whether $f(T)|_{T_i=T_e}$ dependence is mainly due to the change of $T_i$ or $T_e$.

Now we can define the volumetric mass-specific heat capacity without the kinetic contribution of ions ${[\mathcal{C}_v-\mathcal{C}_{v~i}^\mathrm{kin}]}$ for the one-temperature case $T_i=T_e=T$:
\begin{equation}
[\mathcal{C}_v-\mathcal{C}_{v~i}^\mathrm{kin}]=\left(\frac{\partial[\mathcal{E}-\mathcal{E}_i^\mathrm{kin}]}{\partial T}\right)_{T_i=T_e}.
\end{equation}
Then Eq.~(\ref{Eq:Derivatives}) for $f=[\mathcal{E}-\mathcal{E}_i^\mathrm{kin}]$ may be written as:
\begin{multline}
[\mathcal{C}_v-\mathcal{C}_{v~i}^\mathrm{kin}]=\bigg[\left(\frac{\partial [\mathcal{E}-\mathcal{E}_i^\mathrm{kin}]}{\partial T_e}\right)_{T_i}+
\\
+\left(\frac{\partial [\mathcal{E}-\mathcal{E}_i^\mathrm{kin}]}{\partial T_i}\right)_{T_e}\bigg]_{T_i=T_e}.
\label{Eq:DerivativesE}
\end{multline}

\begin{figure}
a)~\includegraphics[width=0.95\columnwidth]{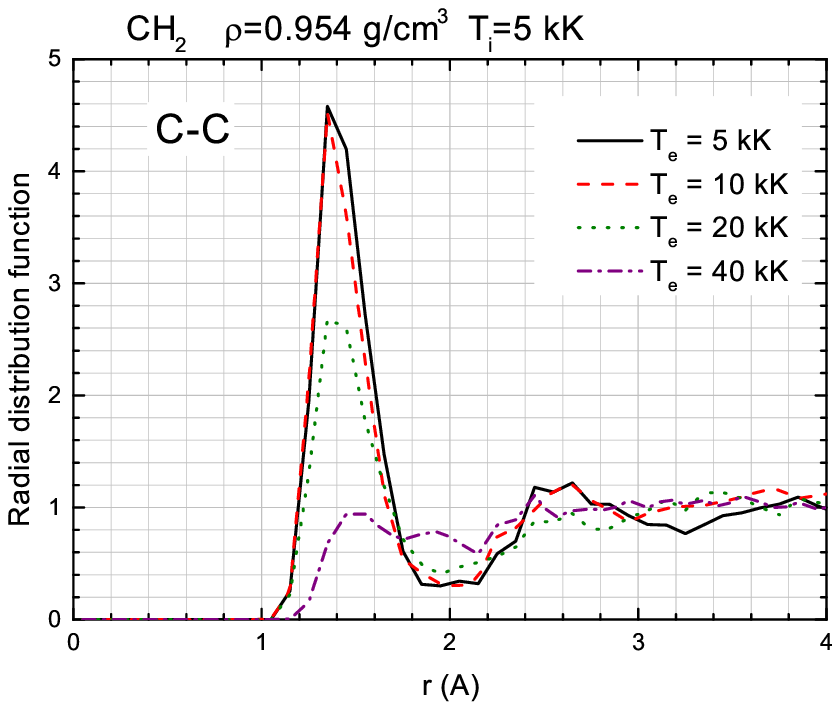}
b)~\includegraphics[width=0.95\columnwidth]{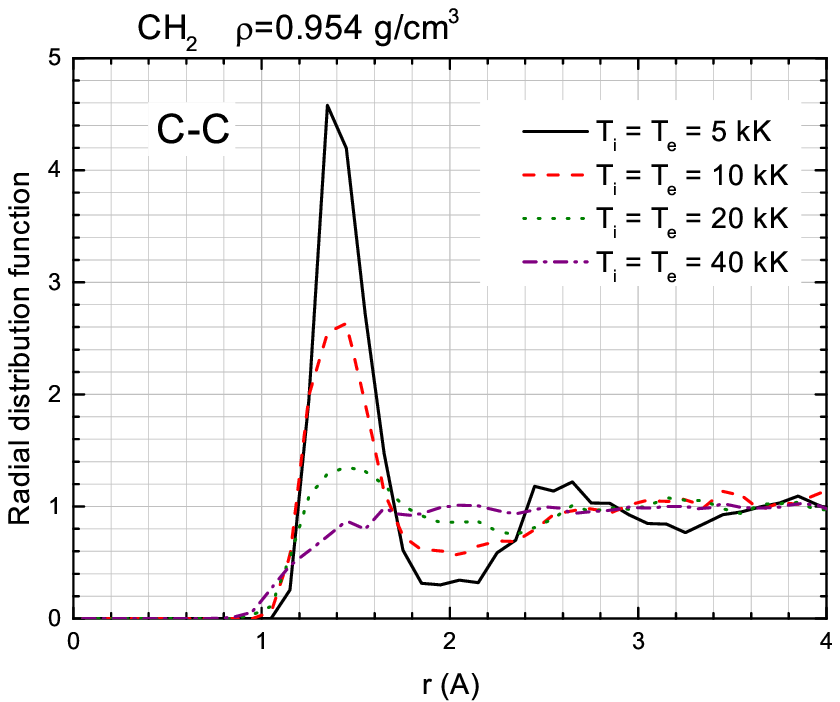}
\caption{C-C radial distribution functions at various temperatures. (a) Two-temperature case: $T_i$ is kept fixed at 5~kK; $T_e$ is varied. (b) One-temperature case: $T_i$ and $T_e$ are varied simultaneously, so that $T_i=T_e=T$. Lines of different types correspond to different temperatures.}
\label{Fig:rdfcc}
\end{figure}

\begin{figure}
a)~\includegraphics[width=0.95\columnwidth]{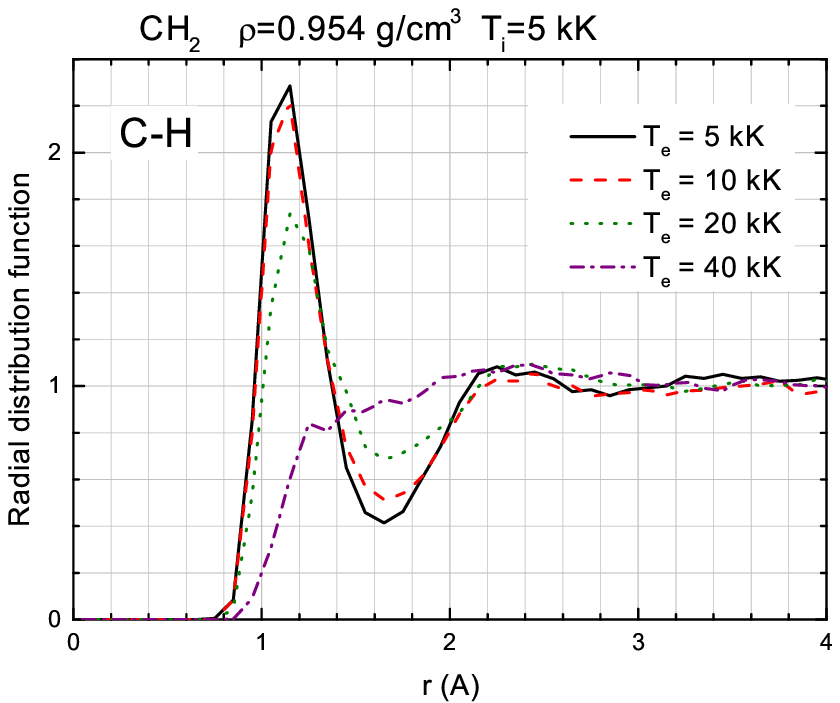}
b)~\includegraphics[width=0.95\columnwidth]{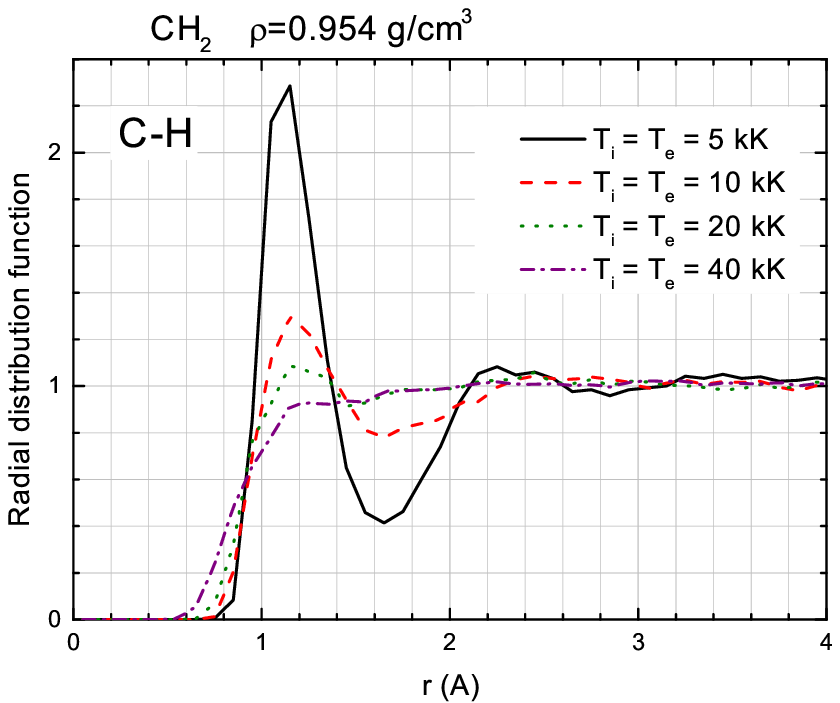}
\caption{Same as Fig.~\ref{Fig:rdfcc} but for C-H radial distribution functions.}
\label{Fig:rdfch}
\end{figure}

\begin{figure}
a)~\includegraphics[width=0.95\columnwidth]{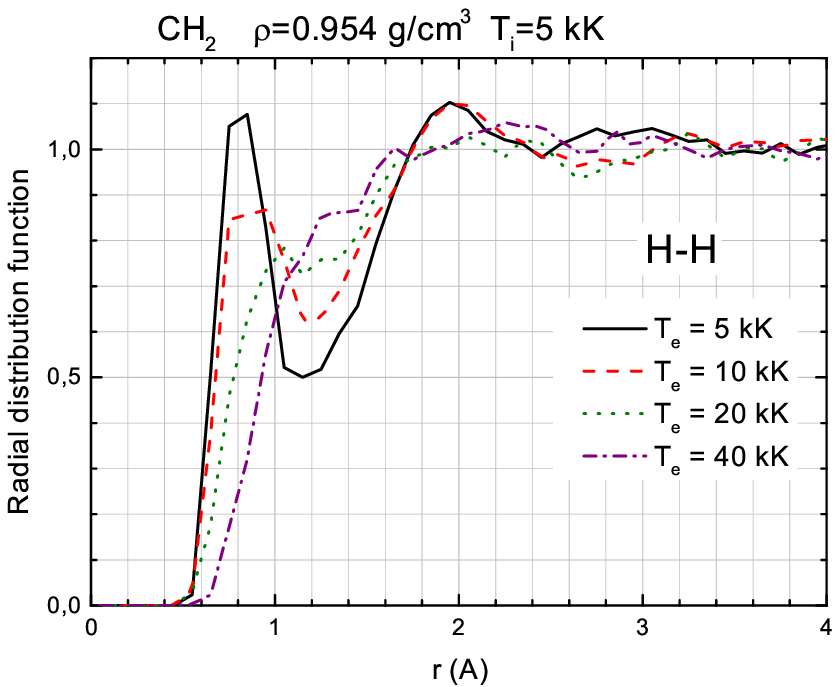}
b)~\includegraphics[width=0.95\columnwidth]{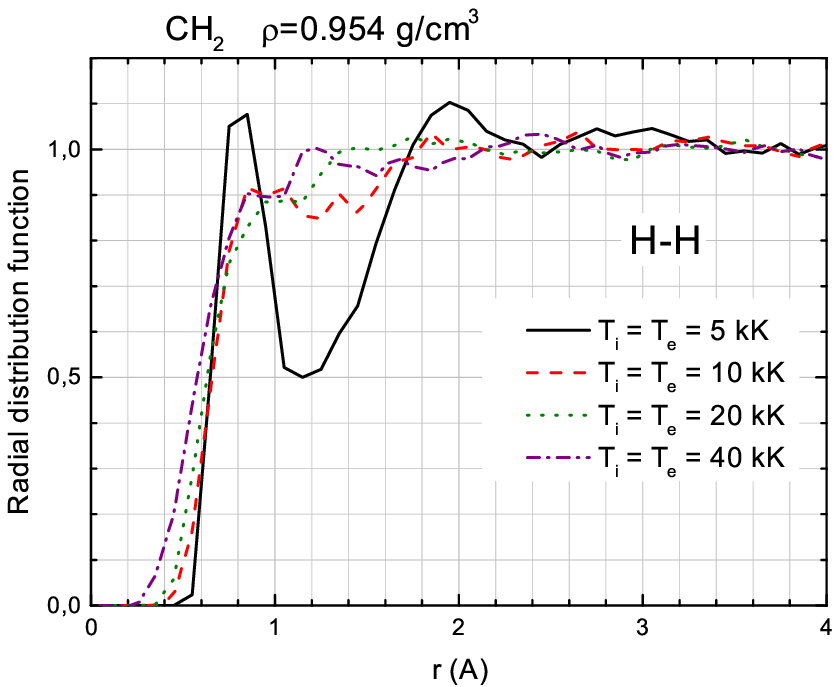}
\caption{Same as Fig.~\ref{Fig:rdfcc} but for H-H radial distribution functions.}
\label{Fig:rdfhh}
\end{figure}

Radial distribution functions were calculated for two cases only:
\begin{itemize}[label={--},noitemsep,nolistsep]
\item one-temperature case 5~kK~$\leq T_i=T_e=T\leq40$~kK (Fig. \ref{Fig:rdfcc}(b), Fig.~\ref{Fig:rdfch}(b), Fig.~\ref{Fig:rdfhh}(b)); 
\item two-temperature case $T_i=5$~kK; 5~kK~$\leq T_e\leq40$~kK (Fig.~\ref{Fig:rdfcc}(a), Fig.~\ref{Fig:rdfch}(a), Fig.~\ref{Fig:rdfhh}(a)).
\end{itemize}
Carbon-carbon (Fig.~\ref{Fig:rdfcc}), carbon-hydrogen (Fig.~\ref{Fig:rdfch}) and hydrogen-hydrogen (Fig.~\ref{Fig:rdfhh}) RDFs are presented.

It is convenient to start the discussion of the results from the RDFs. The general view in Figs.~\ref{Fig:rdfcc}--\ref{Fig:rdfhh} shows, that there are peaks at the RDF curves at low temperatures; these peaks vanish at higher temperatures. In the further discussion we will assume that these peaks are due to the chemical bonds.

Strictly speaking, the presence of chemical bonds may not be reliably established based on the analysis of RDFs only. The peaks at RDF curves have only the following meaning: the interionic distances possess in average certain values $\{r_\mathrm{peak}\}$ more often than other values. But nothing may be said about how long the ions are located at these $\{r_\mathrm{peak}\}$ distances from each other. We should check that ions are located at $\{r_\mathrm{peak}\}$ for certain periods of time $\{t_\mathrm{peak}\}$; only in this case we may establish reliably the presence of chemical bonds. These periods $\{t_\mathrm{peak}\}$ may be called the lifetimes of chemical bonds. This complicated analysis may be found in the papers \cite{Mattsson:PRB:2010, Magyar:PRB:2015}. However, in the current work we will use only RDFs to register chemical bonds.

Fig.~\ref{Fig:rdfcc}(a), Fig.~\ref{Fig:rdfch}(a), Fig.~\ref{Fig:rdfhh}(a) show the RDFs at fixed $T_i=5$~kK and various $T_e$ from 5~kK up to 40~kK. If $T_e$ increases from 5~kK to 10~kK, the bonds are almost intact and the RDF curves almost do not change (only H-H bonds decay to some extent). At $T_e = 20$~kK C-C and C-H bonds are mostly intact (though the peaks become lower); only H-H bonds break almost totally. And only if $T_e$ is risen to 40 kK all the bonds decay.

The situation is different in the one-temperature case $T_i=T_e=T$ (Fig.~\ref{Fig:rdfcc}(b), Fig.~\ref{Fig:rdfch}(b), Fig.~\ref{Fig:rdfhh}(b)). The increase of $T$ from 5~kK to 10~kK already makes the bonds decay. Given the influence of $T_e$ on the bonds is rather weak in this temperature range (see above), this breakdown of bonds is totally due to the increase of $T_i$. At $T=10$~kK H-H bonds are already destroyed totally (Fig.~\ref{Fig:rdfhh}(b)), C-H bonds---almost totally (Fig.~\ref{Fig:rdfch}(b)), \mbox{C-C} bonds---considerably (Fig.~\ref{Fig:rdfcc}(b)). The increase of $T$ to 20~kK leads to the almost complete decay of all bonds.

The following conclusions may be derived from the performed consideration of the RDF curves. If $T_i$ is kept rather low (5~kK), $T_e$ should be risen to 20~kK--40~kK to destroy the bonds. If both $T_e$ and $T_i$ are increased simultaneously (and $T_e=T_i$), temperatures 10~kK--20~kK are quite enough to break the bonds.

\begin{figure}
a)~\includegraphics[width=0.95\columnwidth]{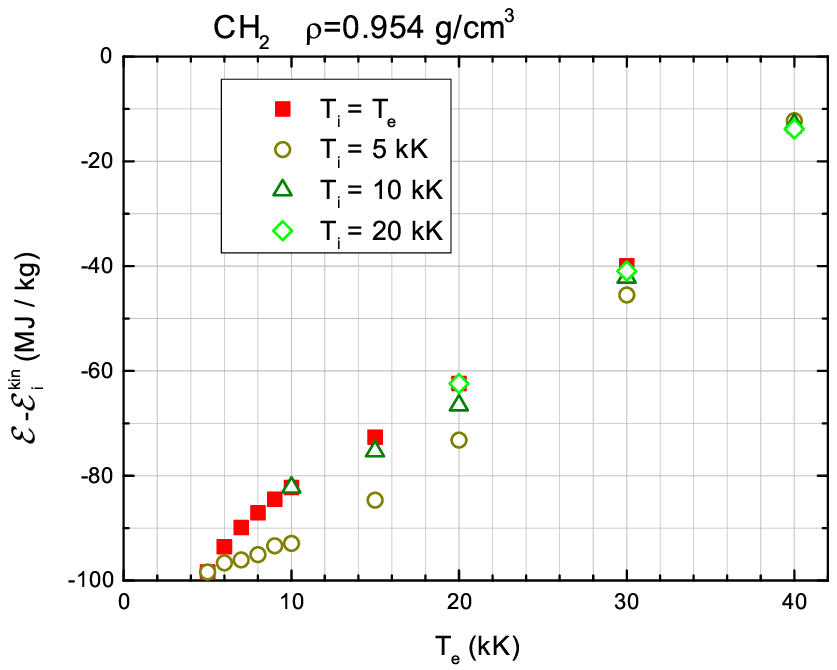}
b)~\includegraphics[width=0.95\columnwidth]{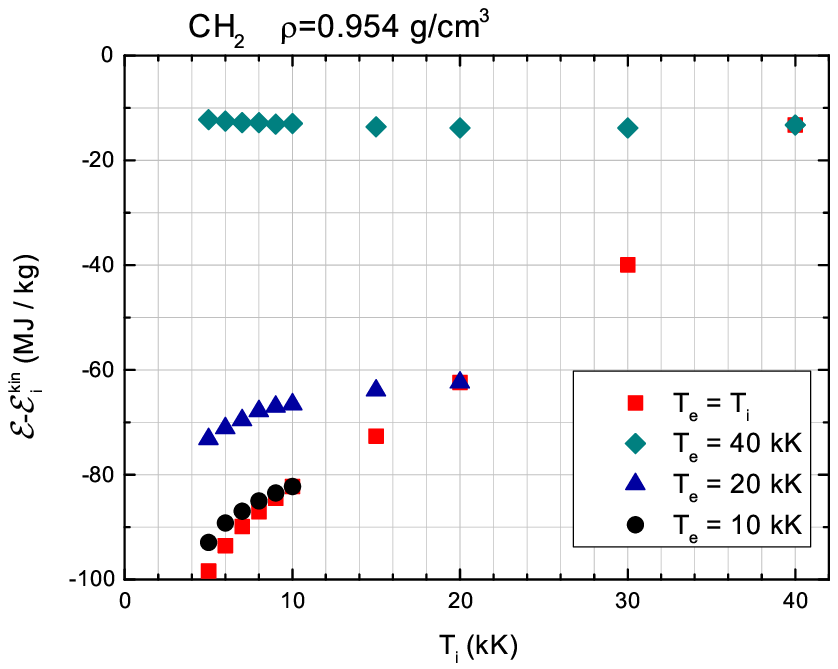}
\caption{Specific energy without kinetic contribution of ions $[\mathcal{E}-\mathcal{E}_i^\mathrm{kin}]$ at different temperatures. (a) Dependence on the electron temperature $T_e$. Filled squares---one-temperature case $T_i=T_e$. Empty circles---fixed $T_i=5$~kK; empty triangles---fixed $T_i=10$~kK; empty diamonds---fixed $T_i=20$~kK. (b) Dependence on the ion temperature $T_i$. Filled squares---one-temperature case $T_e=T_i$. Filled circles---fixed $T_e=10$~kK; filled triangles---fixed $T_e=20$~kK; filled diamonds---fixed $T_e=40$~kK.}
\label{Fig:energynokin}
\end{figure}

\begin{figure}
a)~\includegraphics[width=0.95\columnwidth]{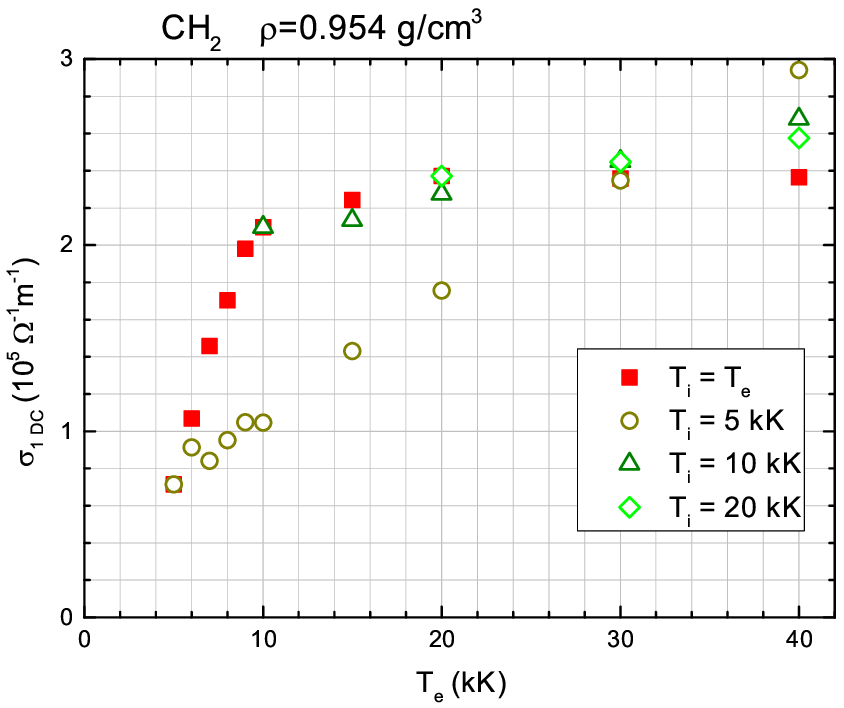}
b)~\includegraphics[width=0.95\columnwidth]{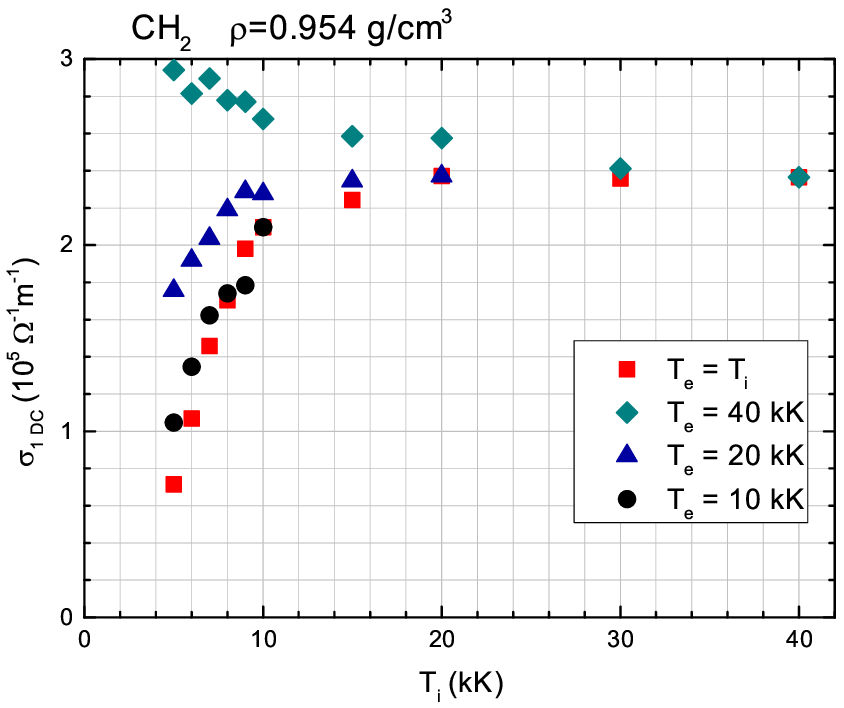}
\caption{Same as Fig.~\ref{Fig:energynokin}, but for static electrical conductivity $\sigma_{1_\mathrm{DC}}$.}
\label{Fig:resigmadc}
\end{figure}

The temperature dependences of $[\mathcal{E}-\mathcal{E}_i^\mathrm{kin}]$ and $\sigma_{1_\mathrm{DC}}$ are presented in Figs.~\ref{Fig:energynokin}--\ref{Fig:resigmadc}. The behavior of the calculated properties depends largely on whether the bonds are destroyed or not.

The obtained results may be divided into three characteristic cases.

1) 5~kK~$\leq T_i\leq T_e\leq10$~kK. The considerable number of bonds are present in the system under these conditions. The chemical bonds break rapidly as $T_i$ grows, and decay rather slowly as $T_e$ increases.

$[\mathcal{E}-\mathcal{E}_i^\mathrm{kin}]$ increases rapidly as $T_i$ grows (Fig.~\ref{Fig:energynokin}(b)), and increases slowly as $T_e$ grows (Fig.~\ref{Fig:energynokin}(a)). $\sigma_{1_\mathrm{DC}}$ increases rapidly as $T_i$ grows (Fig.~\ref{Fig:resigmadc}(b)) and increases slowly as $T_e$ grows (Fig.~\ref{Fig:resigmadc}(a)).

Thus the growth of $[\mathcal{E}-\mathcal{E}_i^\mathrm{kin}]$ and the growth of $\sigma_{1_\mathrm{DC}}$ correlate somewhat with the destruction of the chemical bonds: these processes occur rapidly if $T_i$ rises, and slowly with the rise of $T_e$.

In the one-temperature situation (5~kK~$\leq T_i=T_e=T\leq10$~kK) $[\mathcal{E}-\mathcal{E}_i^\mathrm{kin}](T)$ increases as $T$ grows, this increase is mostly determined by $T_i$ influence (i.e. by the second term in the right hand side of Eq.~(\ref{Eq:Derivatives})). The $\left(\frac{\partial[\mathcal{E}-\mathcal{E}_i^\mathrm{kin}]}{\partial T_i}\right)_{T_e}$ contribution to $[\mathcal{C}_v-\mathcal{C}_{v~i}^\mathrm{kin}]$ is larger than $\left(\frac{\partial[\mathcal{E}-\mathcal{E}_i^\mathrm{kin}]}{\partial T_e}\right)_{T_i}$ (see Eq.~(\ref{Eq:DerivativesE}) and Fig.~\ref{Fig:energynokin}). Since the growth of $[\mathcal{E}-\mathcal{E}_i^\mathrm{kin}]$ correlates with the destruction of bonds here, we may assume that the energy supply necessary for the bond decay gives the main contribution to $[\mathcal{C}_v-\mathcal{C}_{v~i}^\mathrm{kin}]$.

The rapid growth of $\sigma_{1_\mathrm{DC}}(T)$ in the one-temperature situation is mostly due to the influence of $T_i$.

2) 20~kK~$\leq T_i\leq T_e\leq40$~kK. There are no chemical bonds in the system under these conditions.

$[\mathcal{E}-\mathcal{E}_i^\mathrm{kin}]$ grows as $T_e$ increases (Fig.~\ref{Fig:energynokin}(a)) and is almost independent of $T_i$ (Fig.~\ref{Fig:energynokin}(b)). $[\mathcal{E}-\mathcal{E}_i^\mathrm{kin}]$ includes the kinetic energy of electrons, electron-electron, electron-ion and ion-ion potential energies. The fact, that $[\mathcal{E}-\mathcal{E}_i^\mathrm{kin}]$ does not depend on $T_i$, is intuitively clear: there are no significant changes of the ionic structure under the conditions considered (the decay of chemical bonds could be mentioned as a possible example of such changes).

$\sigma_{1_\mathrm{DC}}$ increases as $T_e$ grows (Fig.~\ref{Fig:resigmadc}(a)) and decreases as $T_i$ grows (Fig.~\ref{Fig:resigmadc}(b)).

In the one-temperature situation (20~kK~$\leq T_i=T_e=T\leq40$~kK) $[\mathcal{E}-\mathcal{E}_i^\mathrm{kin}](T)$ increases as $T$ grows only due to $T_e$ influence (i.e. due to the first term in the right hand side of Eq.~(\ref{Eq:Derivatives})). $[\mathcal{C}_v-\mathcal{C}_{v~i}^\mathrm{kin}]$ totally equals $\left(\frac{\partial[\mathcal{E}-\mathcal{E}_i^\mathrm{kin}]}{\partial T_e}\right)_{T_i}$ here (Eq.~(\ref{Eq:DerivativesE})). We can assume that the temperature excitation of the electron subsystem determines $[\mathcal{C}_v-\mathcal{C}_{v~i}^\mathrm{kin}]$ values in this situation.

$\sigma_{1_\mathrm{DC}}$ decreases as $T_i$ grows and increases as $T_e$ grows. In the one-temperature case these two opposite effects compensate each other totally and form $\sigma_{1_\mathrm{DC}}(T)$, that does not depend on $T$.

3) 5~kK~$\leq T_i\leq10$~kK, 30~kK~$\leq T_e\leq40$~kK. This case is qualitatively close to the second one. There are no chemical bonds in the system.

$[\mathcal{E}-\mathcal{E}_i^\mathrm{kin}]$ increases as $T_e$ grows (Fig.~\ref{Fig:energynokin}(a)) and does not depend on $T_i$ (Fig.~\ref{Fig:energynokin}(b)); $\sigma_{1_\mathrm{DC}}$ increases as $T_e$ grows (Fig.~\ref{Fig:resigmadc}(a)) and decreases as $T_i$ grows (Fig.~\ref{Fig:resigmadc}(b)).

\section{Conclusion}

In this paper we have calculated the properties of CH$_2$ plasma in the two-temperature case. First of all, the properties at $T_e> T_i$ are of significant interest for the simulation of rapid laser experiments. The performed calculations also help us to understand better the properties in the one-temperature case $T_i=T_e=T$. Two characteristic regions of the one-temperature curves \cite{Knyazev:PhysPlasmas:2015} may be considered.

The first region corresponds to the temperatures of 5~kK~$\leq T \leq 10$~kK. The significant number of chemical bonds exist in the system in this case. These bonds decay if $T$ is increased (mainly because of heating of ions). We assume, that the energy necessary for the destruction of bonds gives the main contribution to $[\mathcal{C}_v-\mathcal{C}_{v~i}^\mathrm{kin}]$ in this region. The decay of bonds also correlates with the rapid growth of $\sigma_{1_\mathrm{DC}}$.

The second region corresponds to the temperatures of 20~kK~$\leq T\leq 40$~kK. The system contains no chemical bonds under these conditions. The growth of $[\mathcal{E}-\mathcal{E}_i^\mathrm{kin}](T)$ is totally determined by heating of electrons. We assume, that the temperature excitation of the electron subsystem determines $[\mathcal{C}_v-\mathcal{C}_{v~i}^\mathrm{kin}]$ values here. $\sigma_{1_\mathrm{DC}}$ is influenced by heating of both electrons and ions moderately and oppositely. These opposite effects form the plateau on $\sigma_{1_\mathrm{DC}}(T)$ dependence in the second region.

\section*{Acknowledgement}
The majority of computations, development of codes, and treatment of results were carried out in the Joint Institute for High Temperatures RAS under financial support of the Russian Science Foundation (Grant No. 16-19-10700). Some numerical calculations were performed free of charge on supercomputers of Moscow Institute of Physics and Technology and Tomsk State University.


%

\end{document}